\documentclass[final,1p,times]{elsarticle} 
\usepackage{graphicx} 
\usepackage{amssymb} 
\usepackage{amsthm} 
\usepackage{lineno} 

\journal{Nuclear Physics A} 

\let\Otemize =\itemize
\let\Onumerate =\enumerate
\let\Oescription =\description
\def\Nospacing{\itemsep=0pt\topsep=0pt\partopsep=0pt\parskip=0pt\parsep=0pt}
\def\Topspac{\vspace{-0.5\baselineskip}}
\def\Botspac{\vspace{-0.2\baselineskip}}
\newenvironment{Itemize}{\Topspac\Otemize\Nospacing}{\endlist\Botspac}

\begin{document} 

\newcommand{\lsim}{\,{\buildrel < \over {_\sim}}\,}
\newcommand{\gsim}{\,{\buildrel > \over {_\sim}}\,}
\newcommand{\sqrtsNN}{\sqrt{s_{\scriptscriptstyle{{\rm NN}}}}}
\newcommand{\av}[1]{\left\langle #1 \right\rangle}
\newcommand{\eV}{\mathrm{eV}}
\newcommand{\kev}{\mathrm{keV}}
\newcommand{\mev}{\mathrm{MeV}}
\newcommand{\gev}{\mathrm{GeV}}
\newcommand{\tev}{\mathrm{TeV}}
\newcommand{\fm}{\mathrm{fm}}
\newcommand{\mm}{\mathrm{mm}}
\newcommand{\cm}{\mathrm{cm}}
\newcommand{\m}{\mathrm{m}}
\newcommand{\mum}{\mathrm{\mu m}}
\newcommand{\RAA}{R_{\rm AA}}
\newcommand{\RDh}{R_{{\rm D}/h}}
\newcommand{\pt}{p_{\rm t}}
\renewcommand{\d}{{\rm d}}
\newcommand{\dEdx}{{\rm d}E/{\rm d}x}
\newcommand{\dNdy}{{\rm d}N_{\rm ch}/{\rm d}y}
\newcommand{\dNdeta}{{\rm d}N_{\rm ch}/{\rm d}\eta}
\newcommand{\qqbar}{\mbox{$\mathrm {q\overline{q}}$}}
\newcommand{\QQbar}{\mbox{$\mathrm {Q\overline{Q}}$}}
\newcommand{\ppbar}{\mbox{$\mathrm {p\overline{p}}$}}
\newcommand{\ccbar}{\mbox{$\mathrm {c\overline{c}}$}}
\newcommand{\bbbar}{\mbox{$\mathrm {b\overline{b}}$}}
\newcommand{\Dz}{\mbox{$\mathrm {D^0}$}}
\newcommand{\DtoKpi}{\mbox{${\rm D^0\to K^-\pi^+}$}}
\newcommand{\Jpsi} {\mbox{J\kern-0.05em /\kern-0.05em$\psi$}\xspace}

\begin{frontmatter} 


\title{Preparation for heavy-flavour measurements with ALICE\\ at the LHC}

\author{Andrea Dainese, for the ALICE Collaboration}

\address{INFN - Laboratori Nazionali di Legnaro, Legnaro (Padova), Italy}

\begin{abstract} 
ALICE~\cite{aliceJINST} will study nucleus--nucleus and 
proton--proton collisions at the LHC.
The main goal of the experiment is to investigate the properties of QCD matter at 
the extreme energy densities that will be reached in Pb--Pb collisions. 
Heavy quarks (charm and beauty) are regarded as powerful tools for this study.
After briefly reviewing the ALICE heavy-flavour program, we will describe the 
preparation for the first measurements to be performed with pp collisions.
\end{abstract} 

\end{frontmatter} 



\section{Introduction: heavy quarks at the LHC}
\label{intro}

The measurement of open charm and beauty production in 
Pb--Pb collisions at $\sqrtsNN=5.5~\tev$ will allow to 
investigate the mechanisms of heavy-quark production, propagation and
 ha\-dro\-ni\-za\-tion
in the hot and dense QCD 
medium formed in high-energy heavy-ion collisions.
Heavy-quark production measurements in pp collisions 
at  $\sqrt{s}=10$--$14~\tev$, 
besides providing the necessary baseline for
the study of medium effects in Pb--Pb collisions, are interesting 
{\it per se}, as a test of QCD in a new domain, 5--7 times
above the present energy frontier at the Tevatron.


The
$\ccbar$ and $\bbbar$ production yields assumed as the baseline for ALICE simulation studies are: for pp collisions at 
$14~\tev$, 0.16 and 0.007, 
respectively~\cite{alicePPR2} (and lower
by about 25\% at 10~TeV, the envisaged energy of the first long
pp run);  
for the 5\% most central Pb--Pb collisions at $5.5~\tev$,
115 and 4.6, respectively.
These numbers 
are obtained from pQCD calculations at NLO~\cite{hvqmnr} with a reasonable
set of parameters~\cite{alicePPR2}, including 
 nuclear shadowing.
An illustration of 
the theoretical uncertainty bands, spanning over a factor about 2, 
for the D and B 
meson cross sections
will be shown in Section~\ref{central}, 
along with the expected ALICE
sensitivity.

Heavy quark medium-induced quenching is one of the most captivating 
topics to be 
addressed in \mbox{Pb--Pb} collisions at the LHC. Due to the 
QCD nature of parton energy loss, quarks are predicted to lose less
energy than gluons (that have a higher colour charge) and, in addition, 
the `dead-cone effect' is expected to reduce the energy loss of massive 
quarks with respect to light partons~\cite{dk}. 
Therefore, one should observe a pattern 
of decreasing suppression when going from the mostly 
gluon-originated
light-flavour hadrons ($h^\pm$ or $\pi^0$) 
to D and B mesons. In terms of the Pb--Pb-to-pp nuclear modification factors
of the $\pt$-differential yields: 
$\RAA^h(\pt)\lsim\RAA^{\rm D}(\pt)\lsim\RAA^{\rm B}(\pt)$~\cite{adsw}.

\section{Heavy-flavour measurements in preparation}
\label{exp}

The ALICE experimental setup, described in detail in~\cite{aliceJINST,kujer},
allows for the detection of open charm and beauty hadrons
in the high-multiplicity environment 
of central Pb--Pb collisions at LHC energy, where a few thousand 
charged particles might be produced per unit of rapidity. 
The heavy-flavour capability of the ALICE detector is provided by:
\begin{Itemize}
\item Tracking system; the Inner Tracking System (ITS)~\cite{prino}, 
the Time Projection Chamber (TPC)~\cite{jens} and the Transition Radiation Detector (TRD)~\cite{trd},
embedded in a magnetic field of $0.5$~T, allow for track reconstruction in 
the pseudorapidity range $|\eta|<0.9$ 
with a $\pt$ resolution better than
2\% up to $20~\gev/c$ 
and a transverse impact parameter\footnote{
$d_0$, defined as the track distance of closest approach to the 
interaction vertex, in the plane transverse to the beams.} 
resolution better than 
$60~\mum$ for $\pt>1~\gev/c$ 
(the two innermost layers of the ITS, $r\approx 4$ and $7~\cm$, 
are equipped with silicon pixel 
detectors).
\item Particle identification system; charged hadrons are separated via 
$\dEdx$ in the TPC and via time-of-flight in the 
TOF detector; electrons are separated from charged 
hadrons in the dedicated TRD, in the TPC, and in the electromagnetic 
calorimeter (EMCAL); 
muons are identified in the muon 
spectrometer covering the pseudo-rapidity range $-4<\eta<-2.5$. 
\end{Itemize}

Simulation studies~\cite{alicePPR2}
have shown that ALICE has large potential to carry out
a rich heavy-flavour physics programme. The main analyses in preparation 
are:
\begin{Itemize}
\item Open charm: fully reconstructed hadronic decays 
$\rm D^0 \to K^-\pi^+$, $\rm D^+ \to K^-\pi^+\pi^+$, $\rm D^{*+}\to D^0\pi^+$,
$\rm D^0 \to K^-\pi^+\pi^-\pi^+$,
$\rm D_s^+ \to K^-K^+\pi^+$, $\rm \Lambda_c^+ \to p K^-\pi^+$ (under study) in $|\eta|<0.9$; single muons and di-muons in $-4<\eta<-2.5$.
\item Open beauty: 
inclusive single leptons ${\rm B\to e}+X$ 
in $|\eta|<0.9$ and ${\rm B\to\mu}+X$ in $-4<\eta<-2.5$; inclusive displaced
charmonia ${\rm B\to J/\psi\,(\to e^+e^-)}+X$ (under study); b-tagging of 
jets reconstructed in the tracking detectors and in the EMCAL (under study).
\end{Itemize} 

\subsection{Commissioning of the hardware and software tools}
\label{commissioning}

At present, the installation of most of the ALICE detector is completed
and, since December 2007, the different sub-systems are being commissioned 
and calibrated with cosmic-ray tracks (atmospheric 
muons)~\cite{kujer,prino}.
In view of the heavy-flavour measurements, a crucial part of the commissioning
is represented by the alignment of the ITS, that is 
the determination of the actual position and orientation in space
of its 2198 Silicon sensors. The alignment, which has to reach a precision 
well below $10~\mum$ in order to guarantee a close-to-design tracking
resolution, will be performed using tracks from cosmics and first pp collisions.
The first results~\cite{prino}, 
obtained with cosmics collected during summer 2008, 
indicate that the target precision is within reach.

On the offline software side, 
an intense activity for the preparation and refinement of
all the analysis tools is ongoing. In particular, the analysis model 
using the Grid distributed computing environment is being tested.
As an example, in a recent campaign, more than $10^8$ pp events, 
corresponding to about 1/10 of the expected yearly statistics,
have been simulated by running up to 15,000 simultaneous processes 
at almost 100 centres worldwide. The simulated data are now being analyzed 
remotely by several tens of single users.
In the following Sections, we report on 
a selection of results extrapolated to the 
expected statistics collected by ALICE per LHC year
\footnote{$10^7$ central (0--5\% $\sigma^{\rm inel}$) \mbox{Pb--Pb} events in 
1 month at
$\mathcal{L}_{\rm Pb-Pb}=10^{27}~\cm^{-2}{\rm s}^{-1}$
and $10^9$ pp events in 8 months at 
$\mathcal{L}_{\rm pp}^{\rm ALICE}=5\times 10^{30}~\cm^{-2}{\rm s}^{-1}$,
in the barrel detectors; the forward muon arm will collect
about 40 times larger samples of muon-trigger events
(i.e.\, $4\times 10^8$ central \mbox{Pb--Pb} events); safety factors are included.}.

\subsection{Charm and beauty measurements at central rapidity}
\label{central}

\begin{figure}[!t]
  \begin{center}
  \includegraphics[width=0.42\textwidth]{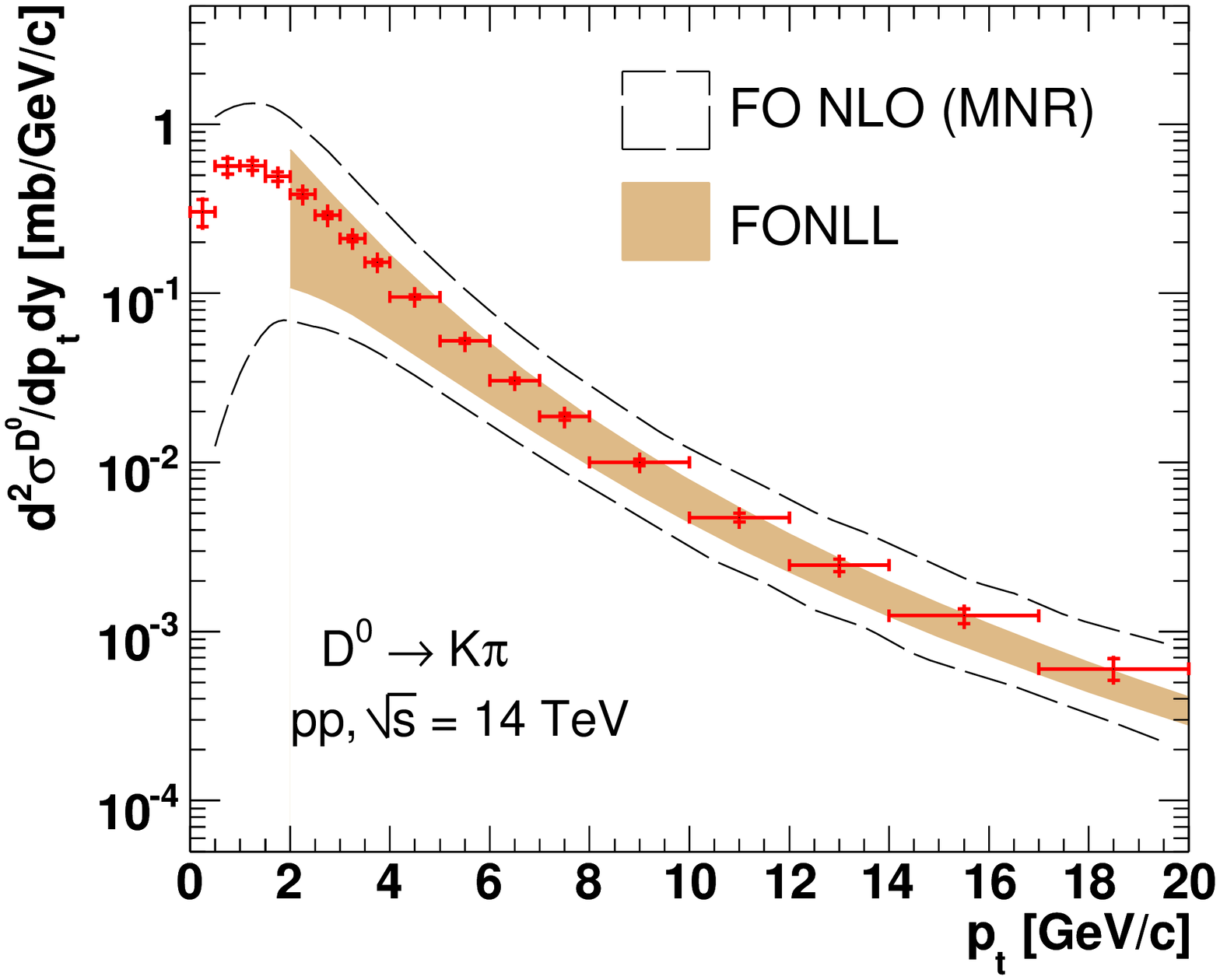}
  \includegraphics[width=0.42\textwidth]{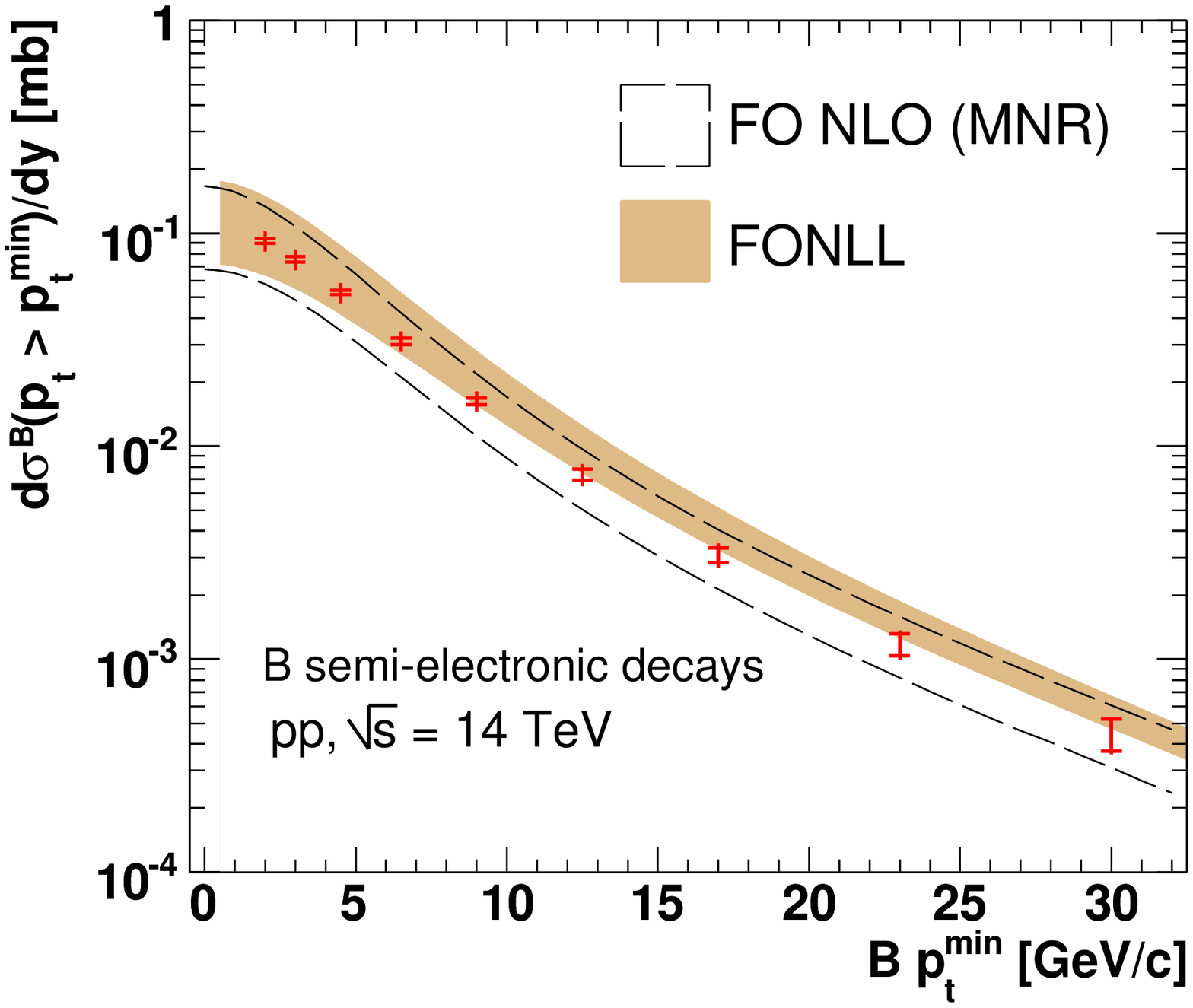}
  \caption{D$^0$ $\pt$-differential (left) and B 
           $\pt^{\rm min}$-differential (right) 
           production cross sections in $|y|<1$,
           in pp at 14~TeV, compared to 
             NLO pQCD predictions (MNR~\cite{hvqmnr} and 
             FONLL~\cite{fonll}).
             Inner error bars represent the statistical errors,
             outer error bars represent the quadratic sum of 
             statistical and systematic errors.
              A normalization error of 5\% is not shown.}
\label{fig:D0Btoe}
\end{center}
\end{figure}

Among the most promising channels for open charm detection are the 
$\rm D^0 \to K^-\pi^+$ ($c\tau\approx 120~\mum$, branching ratio 
$\approx 3.8\%$) and $\rm D^+ \to K^-\pi^+\pi^+$ ($c\tau\approx 300~\mum$, 
branching ratio $\approx 9.2\%$) decays, reconstructed in the 
TPC and ITS, in the rapidity range $|y|<1$. The detection strategy
to cope with the large combinatorial background from the underlying event 
is based on the selection of displaced-vertex topologies, i.e. separation 
of the decay tracks from the primary vertex 
and good alignment between the reconstructed D meson momentum 
and flight direction~\cite{alicePPR2}. 
The accessible $\pt$ range for the $\Dz$ is $1$--$20~\gev/c$ in \mbox{Pb--Pb} and $0.5$--$20~\gev/c$ in pp, 
with statistical errors better than 15--20\% at high $\pt$. Similar capability 
is expected for the $\rm D^+$.
The systematic errors 
(acceptance and efficiency corrections, 
centrality selection for Pb--Pb) are expected to be smaller than 20\%.
The production of open beauty at central rapidity, $|y|<1$, 
can be studied by detecting the 
semi-electronic decays of b-hadrons (branching ratio $\simeq 10\%$). 
Given that electrons from beauty have an average 
impact parameter $d_0\simeq 500~\mum$,
it is possible to 
obtain a high-purity sample with a strategy that relies on
electron identification (TPC and TRD)
and impact parameter cut (to 
reduce the semi-electronic charm-decay component and 
reject misidentified $\pi^\pm$ and $\rm e^{\pm}$
from Dalitz decays and $\gamma$ conversions).
As an example, with $10^7$ central 
\mbox{Pb--Pb} events, this strategy is expected to allow for 
the measurement of the b-decay electron
$\pt$-dif\-fe\-ren\-tial cross section in the range $2<\pt<20~\gev/c$ 
with statistical errors lower
than 15\% at high $\pt$. Similar performance figures are expected for 
pp collisions.
In Fig.~\ref{fig:D0Btoe} we superimpose the simulated results 
for D$^0$ $\d^2\sigma/\d\pt\d y$ and B 
$\d\sigma(\pt>\pt^{\rm min})/\d y$ in pp collisions
to the predictions from the MNR~\cite{hvqmnr} and FONLL~\cite{fonll}
calculations.
The comparison shows that ALICE will be able  to perform 
a sensitive test of the pQCD predictions for c and b production at LHC energy.

By comparing to theoretical predictions 
the expected ALICE precision for the measurement
of the nuclear modification factors $\RAA^{\rm D}$ 
and $\RAA^{\rm e~from~B}$, and for the heavy-to-light ratio  
$\RAA^{\rm e~from~B}/\RAA^{\rm e~from~D}$, 
it has been shown~\cite{DaineseSQM} that the charm and beauty measurements described above can be used
to test the expected colour-charge and mass dependence of parton energy loss.

\subsection{Charm and beauty measurements at forward rapidity}
\label{forward}

\begin{figure}[!t]
  \begin{center}
  \includegraphics[width=0.45\textwidth]{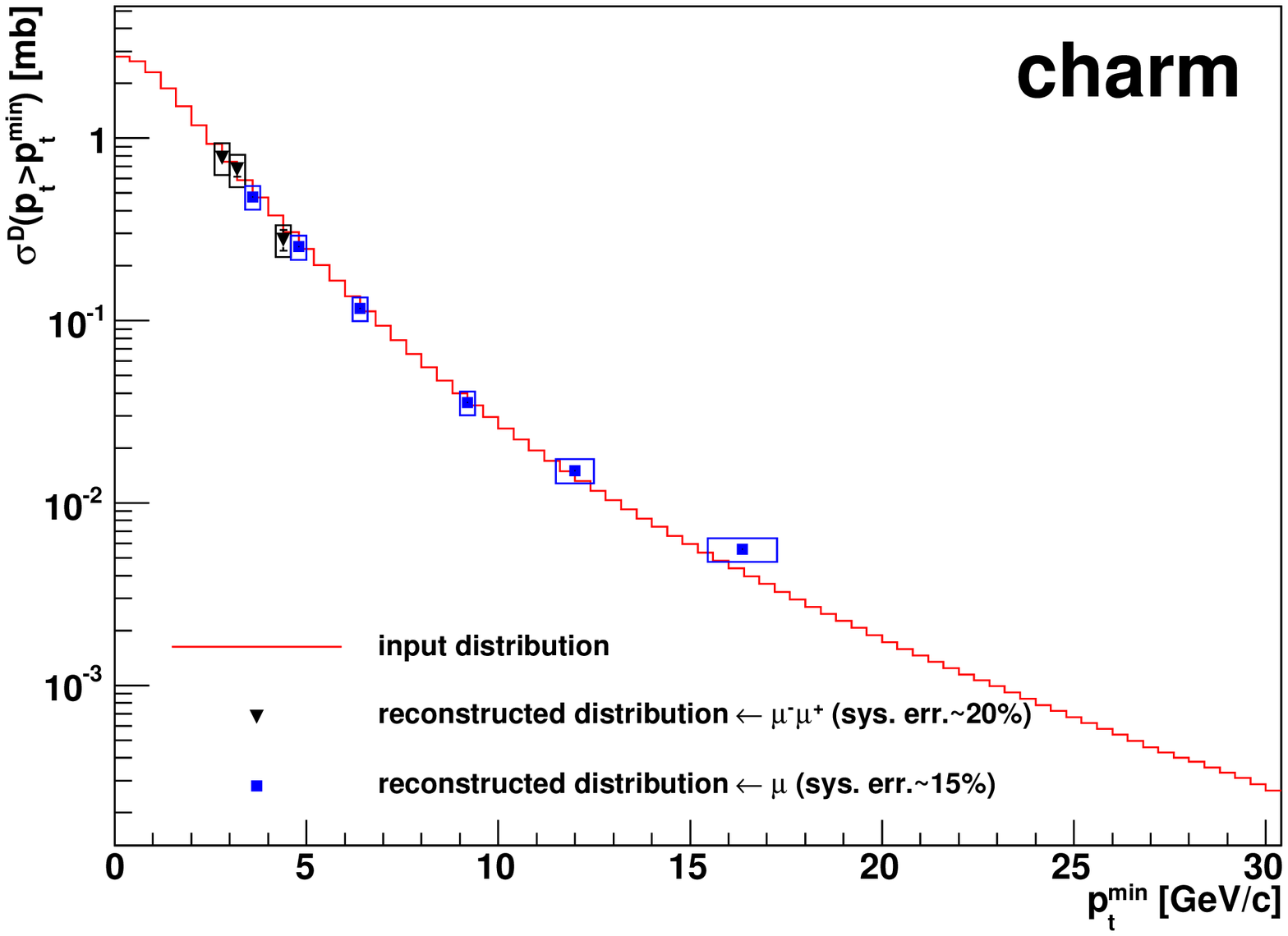}
  \includegraphics[width=0.45\textwidth]{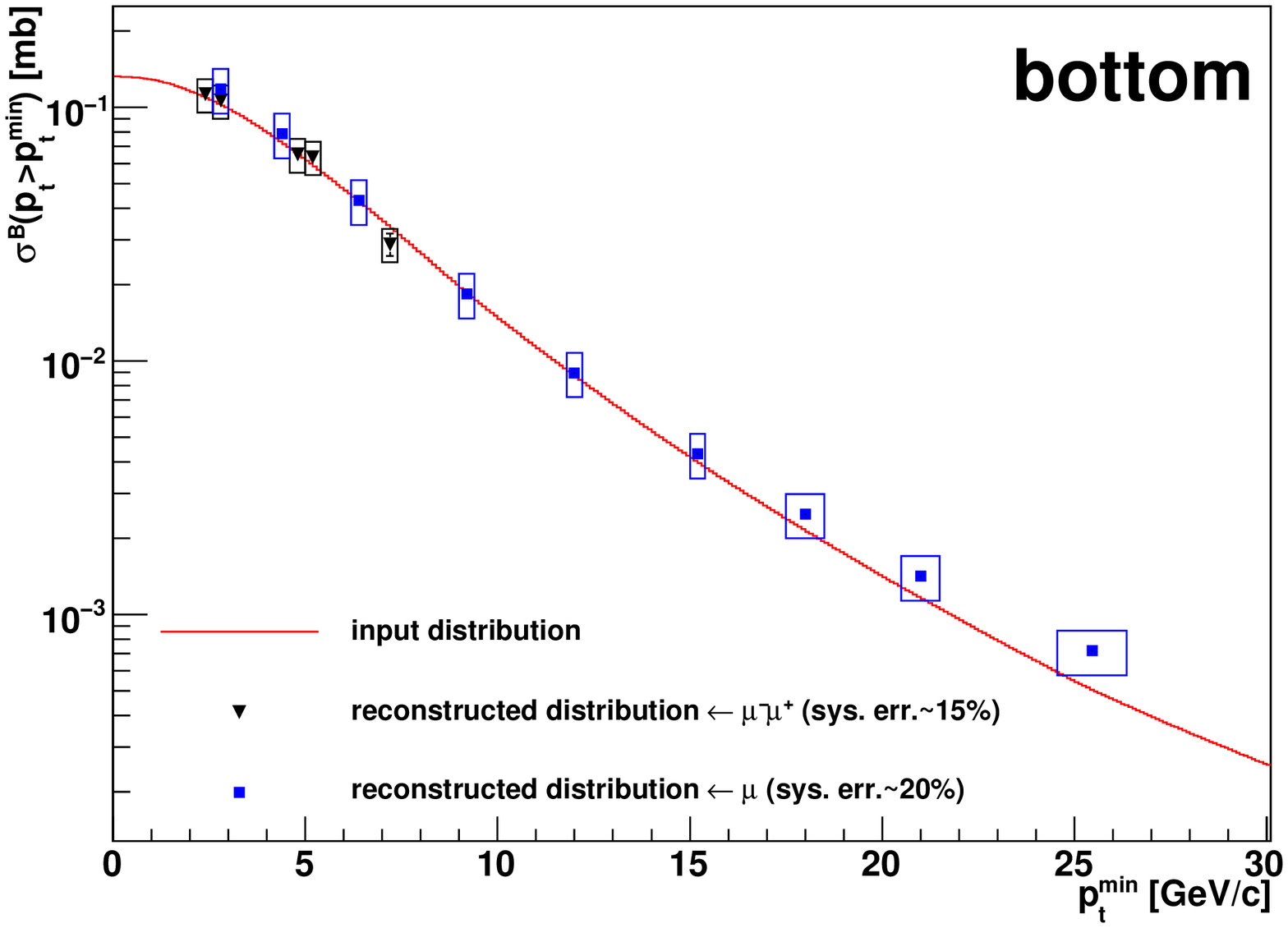}
  \caption{Charm (left) and beauty (right) production measurements
           in $-4<y<-2.5$, using single muons and di-muons, 
           in pp at $\sqrt{s}=14~\tev$. Boxes represent the systematic 
           uncertainties. Error bars represent the
           statistical uncertainties.}
\label{fig:muons}
\end{center}
\end{figure}

Charm and beauty production 
can be measured also in the forward muon 
spectrometer ($-4<\eta<-2.5$) by analyzing the single-muon $\pt$ 
and di-muon invariant-mass distributions~\cite{alicePPR2}.
The main background to the `heavy-flavour muon' signal is $\pi^\pm$ and 
$\rm K^\pm$ decays. The cut $\pt>1.5~\gev/c$ is applied in order to increase the signal-to-background ratio.
Then, a technique that performs a simultaneous fit 
of the single-muon and di-muon distributions with the charm and 
beauty components, using the predicted shapes as templates,
 allows to extract a $\pt^{\rm min}$-differential cross section for 
D and B mesons. The expected performance for pp collisions is shown in 
Fig.~\ref{fig:muons}.
Since only minimal cuts are applied, the statistical errors are 
expected to be lower than 5\% up to muon 
$\pt\approx 30~\gev/c$. The systematic errors, mainly due to the fit assumptions, are expected to be lower than 20\%.
High-$\pt$ single muons could provide the first observation of 
b-quark energy loss at LHC. Indeed,  
the single-muon $\pt$ distribution at LHC energies is expected to 
be dominated by b decays in the range $3\lsim\pt\lsim 25~\gev/c$ and by W-boson 
decays above this range. Therefore, the central-to-peripheral 
muon nuclear modification factor of ${\rm R}_{\rm CP}(\pt)$ would be suppressed 
in the region dominated by beauty, due to parton energy loss, and 
would rapidly increase to about one (binary scaling), where the
medium-blind muons from W decays dominate~\cite{muons}. 

\section{Summary}

Heavy quarks will 
provide ways to test different aspects of 
QCD under extreme conditions at the LHC: from the predictions of 
pQCD at a new energy scale, in pp collisions,
to the mechanism of energy loss in a QCD medium, in Pb--Pb.
The ALICE detectors and analysis tools are being commissioned, with the aim 
of achieving
the excellent tracking, vertexing and particle identification performance 
that will allow to accomplish the rich heavy-flavour physics program.




\begin{thebibliography}{00} 
   

\bibitem{aliceJINST}
 ALICE Collaboration, {\it JINST} {\bf 0803} (2008) S08002.

\bibitem{alicePPR2}
  ALICE Collaboration, Physics Performance Report Vol.~II,
  CERN/LHCC 2005-030
  and {\it J.~Phys.}~{\bf G32} (2006) 1295.

\bibitem{hvqmnr} 
   M.L. Mangano, P. Nason and G. Ridolfi, 
   {\it Nucl. Phys.} {\bf B373} (1992) 295.

\bibitem{dk}
  Yu.L. Dokshitzer and D.E. Kharzeev, {\it Phys.~Lett.} {\bf B519} (2001) 199.

\bibitem{adsw}
  N. Armesto et al., {\it Phys.~Rev.}~{\bf D71} (2005) 054027.

\bibitem{kujer} 
P. Kujer [ALICE Collaboration], {\it these proceedings.}

\bibitem{prino} 
F. Prino [ALICE Collaboration], {\it these proceedings.}

\bibitem{jens} 
J. Wiechula [ALICE Collaboration], {\it these proceedings.}

\bibitem{trd} 
M. Kweon [ALICE Collaboration], {\it these proceedings.}

\bibitem{fonll}
  M. Cacciari et al., {\it JHEP} {\bf 0407} (2004) 033. M. Cacciari, private communication.

\bibitem{DaineseSQM}
  A. Dainese [ALICE Collaboration], {\it J. Phys.} {\bf G35} (2008) 044046;
 {\it Nucl. Phys.} {\bf A783} (2007) 417.

\bibitem{muons}
  Z.~Conesa del Valle et al.,
  {\it Phys.\ Lett.}\  {\bf B663} (2008) 202.
  

\end{thebibliography}
\end{document}